\documentclass[letterpaper,notoc]{CECS}
\title{Quasinormal modes for massless topological \\black holes}
\author{Rodrigo  Aros$^{\dagger,\ddagger}$,  Cristi\'an Mart\'{\i}nez$^\dagger$, 
Ricardo Troncoso$^\dagger$ and Jorge Zanelli$^\dagger$
\footnote{{\it E-mail:} {\tt raros@abello.unab.cl, martinez@cecs.cl, 
ratron@cecs.cl, jz@cecs.cl}}\\$^\dagger$Centro de Estudios Cient\'{\i}ficos (CECS), 
Casilla 1469, Valdivia, Chile
\\$^\ddagger$Universidad Nacional Andr\'{e}s Bello, Sazie 2320, Santiago, Chile  }
\preprint{{\tiny CECS-PHY-02/07}}

\abstract{
An exact expression for the quasinormal modes of  scalar perturbations on a
massless topological black hole in four and higher dimensions is presented.
The massive scalar field is nonminimally coupled to the curvature, and the
horizon geometry is assumed to have a negative constant curvature.
}

\usepackage{amssymb}
\usepackage{amsgen}
\usepackage{amsbsy}

\begin{document}

\section{Introduction}

In a black hole geometry, the event horizon acts as a sink that drains the
linearized perturbations of the geometry or matter fields, damping the
oscillations. These are the so-called quasinormal modes, which are typically
characterized by a spectrum that is independent of the initial conditions.
The quasinormal modes are a sort of fingerprint of the black hole depending
only on its parameters and on the fundamental constants of the system.

Quasinormal modes have been extensively studied in asymptotically flat
spacetimes (see, \emph{e.g}\textit{.}, Ref. \cite{Kokkotas:1999bd} and references
therein). The inclusion of a negative cosmological constant adds a new angle
of interest to the problem [2 - 18].
Through
the AdS/CFT correspondence, the quasinormal modes can be related to the
relaxation time scale of the associated thermal states \cite
{Horowitz:1999jd, Witten:1998zw, Aharony:1999ti}. Recently, a connection has
also been conjectured between the quasinormal modes and critical phenomena
of black hole formation in an asymptotically AdS background \cite
{Horowitz:1999jd}. In three dimensions analytic results supporting these
conjectures have been established recently [5 - 10].
In four dimensions, however, the quasinormal
frequencies on black holes have been obtained by numerical methods only.

A negative cosmological constant allows the existence of black holes with
a topology $\Bbb{R}^{2}\times \Sigma $, where $\Sigma $ is a two-dimensional
manifold of constant curvature \cite{Lemos, Vanzo:1997gw,Brill:1997mf}. The
simplest solution of this kind, when $\Sigma $ has negative constant
curvature, reads
\begin{equation}
ds^{2}=-\left( -1+\frac{r^{2}}{l^{2}}-\frac{2\mu }{r}\right) dt^{2}+\frac{
dr^{2}}{\left( -1+\frac{r^{2}}{l^{2}}-\frac{2\mu }{r}\right) }+r^{2}d\sigma
^{2}\;,  \label{Top-BH-Einstein}
\end{equation}
where the constant $\mu $ is proportional to the mass and is bounded from
below as $\mu \geq -\frac{l}{3\sqrt{3}}$. Here $l$ is the AdS radius, and $%
d\sigma ^{2}$ is the line element of $\Sigma $, which must be locally
isomorphic to the hyperbolic manifold $H^{2}$. By virtue of the Killing-Hopf
theorem, $\Sigma $ must be of the form
\[
\Sigma =H^{2}/\Gamma {\quad }\mathrm{{with\quad }\Gamma \subset O(2,1)\;,}
\]
where $\Gamma $ is a freely acting discrete subgroup (i.e., without fixed
points).

The configurations (\ref{Top-BH-Einstein}) are asymptotically locally AdS
spacetimes. These spacetimes can admit Killing spinors for $\mu =0$ provided
$\Sigma $ is a noncompact surface \cite{Aros:2002rk}. In this case Eq. (\ref
{Top-BH-Einstein}) describes a warped black string with a supersymmetric
ground state for $\mu =0$, and therefore expected to be stable. On the other
hand, it has been recently shown in Ref. \cite{Gibbons:2002pq} that the massless
configurations where $\Sigma $ has negative constant curvature are stable
under gravitational perturbations.

In this paper, the exact expression for the quasinormal modes of a massive
scalar field in the geometry (\ref{Top-BH-Einstein}) with $\mu =0$ is
presented. Since this geometry has constant curvature, the inclusion of a
conformal coupling amounts just to a shift in the mass parameter of the
scalar field. The generalization of the expression for the quasinormal modes in higher
dimensions is also obtained.

\section{Quasinormal modes in four dimensions}

Consider the exterior region of the black hole (\ref{Top-BH-Einstein}) in
the massless case
\[
ds^{2}=-\left( \frac{r^{2}}{l^{2}}-1\right) dt^{2}+\frac{dr^{2}}{\left(
\frac{r^{2}}{l^{2}}-1\right) }+r^{2}d\sigma ^{2}\;.
\]
This is a manifold of negative constant curvature possessing an event
horizon at $r=l$. A massive scalar field with a non minimal coupling
satisfies
\begin{equation}
\left(\square -m^{2}-\frac{\gamma }{6}R \right)\phi =0\;,  \label{ScalarEquation}
\end{equation}
which becomes conformally invariant for $m=0$ and $\gamma =1$. Here $\square
$ stands for the Laplace-Beltrami operator. Since the scalar curvature is $
R=-12l^{-2}$, this equation reduces to
\begin{equation}
(\square -m_{\textrm{{\tiny {eff}}}}^{2})\phi =0\;,  \label{Klein-Gordon}
\end{equation}
with an effective mass given by $m_{\textrm{{\tiny {eff}}}}^{2}=m^{2}-2\gamma
l^{-2}$.

The solution of Eq. (\ref{Klein-Gordon}) can be readily found making the
following coordinate transformation $z=1-l^{2}/r^{2}$ and $t\rightarrow lt$,
so that the metric reads
\begin{equation}
ds^{2}=\frac{l^{2}}{(1-z)}\left[ -zdt^{2}+\frac{dz^{2}}{4z(1-z)}+d\sigma
^{2}\right] \;,  \label{MasslessZ}
\end{equation}
where $0\leq z<1$, and adopting the following ansatz
\begin{equation}
\phi =R(z)e^{-i\omega t}Y(\Sigma )\;.  \label{Scalar-Ansatz}
\end{equation}
Here $Y$ is a normalizable harmonic function on $\Sigma $, i.e. it satisfies
$\nabla ^{2}Y=-QY$, where $\nabla ^{2}$ is Laplace operator on $\Sigma $.
The eigenvalues for the hyperbolic manifold $H^{2}$ are
\begin{equation}
Q=\frac{1}{4}+\xi ^{2}\;,  \label{Q-Xi}
\end{equation}
where $\xi $ is any real number, see, \emph{e.g.}, Ref.  \cite{Terras:1985}. 
Since $\Sigma $ is a quotient of the form $H^{2}/\Gamma $, the spectrum has the
form (\ref{Q-Xi}), but the parameter $\xi $ becomes restricted depending on 
$\Gamma $. Indeed, if $\Sigma $ is a closed manifold the spectrum is
discrete. Note that the zero mode, $Q=0$, is not in the spectrum.

The radial function $R(z)$ satisfies
\begin{equation}
\left[ z(1-z)\partial _{z}^{2}+\left( 1-\frac{z}{2}\right) \partial
_{z}+\left( \frac{\omega ^{2}}{4z}-\frac{Q}{4}-\frac{m_{\textrm{{\tiny {eff}}}
}^{2}l^{2}}{4(1-z)}\right) \right] R(z)=0\;.  \label{Rz}
\end{equation}
Under the decomposition $R(z)=z^{\alpha }(1-z)^{\beta }K(z)$, Eq. (\ref{Rz})
becomes the hypergeometric equation for $K$,
\begin{equation}
z(1-z)K^{\prime \prime }+[c-(1+a+b)z]K^{\prime }-abK=0\;,  \label{HyperGeom}
\end{equation}
provided\footnote{%
Actually $\alpha =\pm \frac{i\omega }{2}$, and without loss of generality
the negative sign can be chosen.}
\begin{eqnarray}
\alpha &=&-\frac{i\omega }{2}\;,  \label{alfa} \\
\beta &=&\beta _{\pm }=\frac{3}{4}\pm \frac{1}{4}\sqrt{9+4m_{\textrm{{\tiny {
eff}}}}^{2}l^{2}}\;.  \label{Beta}
\end{eqnarray}
The solution of Eq. (\ref{HyperGeom}) takes the form
\begin{equation}
K=C_{1}F(a,b,c,z)+C_{2}z^{1-c}F(a-c+1,b-c+1,2-c,z)\;,  \label{K}
\end{equation}
where the coefficients are defined as
\begin{eqnarray}
a &=&-\frac{1}{4}+\alpha +\beta _{\pm }+\frac{i\xi }{2}\;,  \nonumber \\
b &=&-\frac{1}{4}+\alpha +\beta _{\pm }-\frac{i\xi }{2}\;,  \label{abc} \\
c &=&1+2\alpha \;,  \nonumber
\end{eqnarray}
and $c$ cannot be an integer.

On physical grounds, $\phi $ must be restricted to be a purely ingoing wave
at the horizon. Furthermore, since the spacetime is locally AdS, the
energy-momentum flux density at the asymptotic region should vanish.

The behavior of the scalar field near the horizon $(z=0)$ is given by
\[
\phi \sim C_{1}e^{-i\omega (t+\ln (z)/2)}+C_{2}e^{-i\omega (t-\ln (z)/2)}\;.
\]
Then, $\phi $ is purely ingoing at the horizon for $C_{2}=0$, and therefore
the radial function is
\begin{equation}
R(z)=z^{\alpha }(1-z)^{\beta }F(a,b,c,z)\;.  \label{Radial}
\end{equation}
The energy-momentum tensor for the scalar field is given by
\[
T_{\mu \nu }=\partial _{\mu }\phi \partial _{\nu }\phi -\frac{1}{2}g_{\mu
\nu }\partial ^{\alpha }\phi \partial _{\alpha }\phi -\frac{m^{2}}{2}g_{\mu
\nu }\phi ^{2}+\frac{\gamma }{6}[g_{\mu \nu }\square -\nabla _{\mu }\nabla
_{\nu }+G_{\mu \nu }]\phi ^{2}\;,
\]
where $G_{\mu \nu }$ is the Einstein tensor. The current $j^{\mu }=\sqrt{-g}
\xi ^{\nu }T_{\;\nu }^{\mu }$, which is conserved if $\xi ^{\mu }$ is a
Killing vector, allows to define the energy choosing $\xi ^{\mu }=\delta
_{0}^{\mu }$. Thus, the energy-momentum flux density $\sqrt{-g}T_{z0}g^{zz}$
at the asymptotic region vanishes if

\[
\lim_{z\rightarrow 1}\frac{1}{\sqrt{1-z}}\left( \left( 1-\frac{2\gamma }{3}
\right) z\partial _{z}+\frac{\gamma }{3}\frac{1}{(1-z)}\right) \phi
^{2}\rightarrow 0\;.
\]
A detailed analysis (see the Appendix) shows that this last condition is
satisfied only if
\[
m_{\textrm{{\tiny {eff}}}}^{2}\geq -\frac{9}{4l^{2}}\;,
\]
which agrees with the Breitenlohner-Freedman bound for the positivity of
energy in global AdS$_{4}$ \cite{Breitenlohner:jf,Breitenlohner:bm}. In this
case, the expression for the scalar field is given by Eqs. (\ref
{Scalar-Ansatz}) and (\ref{Radial}), with $\beta =\beta _{+}$ as defined in Eq. (
\ref{Beta}). The quasinormal frequencies are determined by the conditions $
a|_{\beta _{+}}=-n$ or $b|_{\beta _{+}}=-n$, with $n=0,1,2...$ in Eq.  (\ref{abc}
) yielding

\begin{equation}
\omega =\pm \xi -i\left( 2n+1+\sqrt{\frac{9}{4}+m_{\textrm{{\tiny {eff}}}
}^{2}l^{2}}\right) \;.
\end{equation}
As shown in the appendix, in analogy with the normal modes for AdS$_{4}$, if
the mass and the coupling constant $\gamma $ satisfy
\begin{equation}
\sqrt{\frac{9}{4}+m_{\textrm{{\tiny {eff}}}}^{2}l^{2}}=\frac{3}{2}-\frac{%
\gamma }{3-2\gamma }\;,  \label{CasoRaro}
\end{equation}
there exist an alternative set of modes for the range of effective mass
\begin{equation}
-\frac{9}{4}<m_{\textrm{{\tiny {eff}}}}^{2}l^{2}<-\frac{5}{4}\;,
\label{RangoRaro}
\end{equation}
for which the scalar field is obtained from Eqs. (\ref{Scalar-Ansatz}), (\ref
{Radial}) but with $\beta =\beta _{-}$. In this case, the quasinormal
frequencies are found analogously by $a|_{\beta _{-}}=-n$ or $b|_{\beta
_{-}}=-n$, so that

\begin{equation}
\omega =\pm \xi -i\left( 2n+1-\sqrt{\frac{9}{4}+
m_{\textrm{{\tiny {eff}}}}^{2}l^{2}}\right) \;.
\end{equation}
Note that a massless scalar field ($m=0$) satisfies the condition (\ref
{CasoRaro}) only for conformal coupling ($\gamma =1$). Actually, there are
two other values of $\gamma $ which satisfy  Eq. (\ref{CasoRaro}) for $m=0$,
namely,  $\gamma =0$, $9/8$. These two roots, however, yield an effective mass
which lies outside the range (\ref{RangoRaro}).

Remarkably, the damping time scale is independent of the parameter $\xi $,
which determines the eigenvalue of the Laplacian in $\Sigma $. This is
contrary to the observation in the Schwarzschild-AdS case, where
surprisingly, the damping time scale increases with the angular momentum of
the mode \cite{Horowitz:1999jd}. In the next section, the generalization 
of these results for higher dimensions is discussed.

\section{Higher dimensions}

Black holes with topologically nontrivial transverse sections of negative
constant curvature exist for $d>4$ dimensions \cite{Birmingham:1998nr,
Cai-Soh, Emparan:1999pm} and also for gravity theories containing higher
powers of the curvature \cite{Crisostomo:2000bb,Aros:2000ij}.
Following the procedure of Refs. \cite{Aros:1999kt} and \cite{Aros:1999id},
the mass can be obtained from a surface integral at infinity. From this, it
can be seen that for all cases, the massless solution is described by a
metric of the same general form as in Eq.(\ref{MasslessZ}), but now $d\sigma
^{2}$ stands for the line element of a $(d-2)$-dimensional surface $\Sigma
_{d-2}$ of negative constant curvature. In this case, the configuration (\ref
{MasslessZ}) is a locally AdS spacetime

Spacetimes of the form (\ref{MasslessZ}) in $d$ dimensions admitting Killing
spinors were classified \cite{Aros:2002rk}, where it was shown that global
supersymmetry can be attained only if $\Sigma _{d-2}$ is a noncompact
surface.

Topological black holes have scalar curvature given by $R=-d(d-1)l^{-2}$ and
therefore the massive scalar field with a non minimal coupling satisfies the
Klein-Gordon equation with an effective mass given by
\begin{equation}
m_{\textrm{{\tiny {eff}}}}^{2}=m^{2}-\gamma \frac{d(d-2)}{4l^{2}}\;.
\label{HighereffectiveMass}
\end{equation}
This equation can be solved for the massless background with the same ansatz
as for the four-dimensional case (\ref{Scalar-Ansatz}), but now, $Y$ is a
harmonic function of finite norm with eigenvalue $-Q$ on $\Sigma _{d-2}$.
Since $\Sigma _{d-2}$ is surface of constant curvature it must be $H^{d-2}$
or a quotient thereof, and hence, the spectrum of the Laplace operator takes
the form \cite{Donnolly}
\begin{equation}
Q=\left( \frac{d-3}{2}\right) ^{2}+\xi ^{2},
\end{equation}
where for $H^{d-2}$ the parameter $\xi $ takes all real values, and upon
identifications, the parameter $\xi $ is generically restricted, becoming a
discrete set if $\Sigma _{d-2}$ is a closed surface.

Now, the radial function $R(z)$ satisfies
\[
\left[ z(1-z)\partial _{z}^{2}+\left( 1+\left( \frac{d-5}{2}\right) z\right)
\partial _{z}+\left( \frac{\omega ^{2}}{4z}-\frac{Q}{4}-\frac{m_{\textrm{%
{\tiny {eff}}}}^{2}l^{2}}{4(1-z)}\right) \right] R(z)=0\;,
\]
and if we choose $R=z^{\alpha }(1-z)^{\beta }K(z)$ the function $K$
satisfies the hypergeometric equation (\ref{HyperGeom}) whose solution is
also given by Eq. (\ref{K}) where hypergeometric parameters are now
\begin{eqnarray}
a &=&-\left( \frac{d-3}{4}\right) +\alpha +\beta _{\pm }+\frac{i\xi }{2}\;,
\nonumber \\
b &=&-\left( \frac{d-3}{4}\right) +\alpha +\beta _{\pm }-\frac{i\xi }{2}\;,
\label{abcHigher} \\
c &=&1+2\alpha \;,  \nonumber
\end{eqnarray}
where $c$ is not an integer and
\begin{eqnarray}
\alpha &=&-\frac{i\omega }{2}  \nonumber \\
\beta &=&\beta _{\pm }=\frac{d-1}{4}\pm \frac{1}{2}\sqrt{\left( \frac{d-1}{2}
\right) ^{2}+m_{\textrm{{\tiny {eff}}}}^{2}l^{2}}  \label{BetaHigher}
\end{eqnarray}

In analogy with the four-dimensional case, requiring $\phi $ to be purely
ingoing at the horizon, fixes the form of the radial function as
\begin{equation}
R(z)=z^{\alpha }(1-z)^{\beta }F(a,b,c,z)\;.  \label{RadialHigher}
\end{equation}

As shown in the appendix, the vanishing of the energy-momentum flux density
implies that if the effective mass satisfies the bound
\begin{equation}
m_{\textrm{{\tiny {eff}}}}^{2}l^{2}\geq -\left( \frac{d-1}{2}\right) ^{2}\;,
\label{MT-Bound}
\end{equation}
the scalar field is given by Eqs. (\ref{Scalar-Ansatz}), (\ref{RadialHigher})
with $\beta =\beta _{+}\,$in (\ref{BetaHigher}). The quasinormal frequencies
are determined by $a|_{\beta _{+}}=-n$ or $b|_{\beta _{+}}=-n$, with $
n=0,1,2,...$, in Eq. (\ref{abcHigher}), which yields

\begin{equation}
\omega =\pm \xi -i\left( 2n+1+\sqrt{\left( \frac{d-1}{2}\right) ^{2}+m_{
\textrm{{\tiny {eff}}}}^{2}l^{2}}\right) \;.  \label{omegamas}
\end{equation}
The bound (\ref{MT-Bound}) coincides with the one obtained by Mezincescu and
Townsend for the normal modes in global AdS$_{d}$ \cite{Mezincescu:ev}.

If the mass and the coupling constant $\gamma $ satisfy the relation
\begin{equation}
\sqrt{\left( \frac{d-1}{2}\right) ^{2}+m_{\textrm{{\tiny {eff}}}}^{2}l^{2}}=
\frac{d-1}{2}-\frac{\gamma }{2}\frac{d-2}{d-1-\gamma (d-2)}
\label{CasoRaroHigher}
\end{equation}
for the range of effective mass given by
\begin{equation}
-\left( \frac{d-1}{2}\right) ^{2}<m_{\textrm{{\tiny {eff}}}}^{2}l^{2}<1-\left(
\frac{d-1}{2}\right) ^{2}\;,  \label{RangoRaroHigher}
\end{equation}
there is another set of modes for which the scalar field is obtained from
Eqs. (\ref{Scalar-Ansatz}, \ref{RadialHigher}) but now for $\beta =\beta
_{-} $. This second set of quasinormal frequencies is given by
\begin{equation}
\omega =\pm \xi -i\left( 2n+1-\sqrt{\left( \frac{d-1}{2}\right) ^{2}+m_{
\textrm{{\tiny {eff}}}}^{2}l^{2}}\right) \;.  \label{omegamenos}
\end{equation}
As it occurs in four dimensions, the massless scalar field ($m=0$) satisfies
the condition (\ref{CasoRaroHigher}) only for conformal coupling ($\gamma =1$
). Also, the damping time scale is independent of the eigenvalue of the
laplacian on $\Sigma $.

\section{Discussion and comments}

The scalar perturbations on the massless black hole geometry (\ref{MasslessZ}
) have been described in four and higher dimensions. As the transverse
section $\Sigma _{d-2}$ is a quotient of $H^{d-2}$, the imaginary part of $%
\omega $ is nonnegative, and therefore stability is always guaranteed. The
expressions for the quasinormal frequencies and eigenfunctions are
explicitly found. To these authors' knowledge, this is the only analytic
result for a black hole in four and higher dimensions.

The advantage of working with analytic expressions is that it is possible to
impose the vanishing of the energy-momentum flux at infinity, which is more
general than requiring the vanishing of $\phi $ at infinity as is usually
done in numerical computations. As can be seen from Eq. (\ref{AsymptR}), the
vanishing of the scalar field at infinity leads to the same modes as those
found here for $m_{\textrm{{\tiny {eff}}}}^{2}\geq 0$. However, for the range $
-\left( \frac{d-1}{2}\right) ^{2}\leq m_{\textrm{{\tiny {eff}}}}^{2}l^{2}<0$,
the field vanishes identically at infinity so that this condition does not
yield information about the modes.

A numerical analysis of the quasinormal modes for topological black holes was done in Refs.
\cite{TopQNnumerical} assuming the eigenvalue of the Laplacian on $\Sigma $ to be $Q=0$. This
value, however, is not in the spectrum of the Laplacian and, according to our result, would give
rise to damping without oscillations, or even to unstable modes for certain values of the
effective mass.

According to the AdS/CFT correspondence, the quasinormal modes are related
to the relaxation time scale of a perturbation in the associated thermal
states at the boundary. In this case, the thermal CFT is defined on $
S^{1}\times \Sigma _{d-2}$, and the characteristic time scale is given by $
\tau =(\mathrm{Im}[\omega ]\mathrm{)^{-1}}$, where $\omega $ can be given by
Eq. (\ref{omegamas}) or (\ref{omegamenos}) . Following Ref. \cite
{Horowitz:1999jd}, it would be interesting to compare our results against
the critical exponents of the formation process for these black holes.

\acknowledgments

The authors are grateful to Jaime Vel\'{a}zquez for helpful remarks. This
work is partially funded by grants No. 1010446, 1010449, 1010450, 1020629,
7010446, 7010450 from FONDECYT and grant No. DI 08-02 (UNAB). 
The generous support of the Empresas CMPC  to the Centro de Estudios
Cient\'{\i}ficos (CECS) is gratefully acknowledged. CECS is a Millennium
Science Institute and is funded in part by grants from  Fundaci\'on Andes and
 the Tinker Foundation.

\appendix

\section{Appendix}

Consider a massive real scalar field in $d$ dimensions nonminimally coupled
to the background geometry
\begin{equation}
\left( \square -m^{2}-\frac{\gamma }{4}\frac{d-2}{d-1}R\right) \phi =0\;,
\end{equation}
which becomes conformally invariant for $\gamma =1$ and $m=0$. The
energy-momentum tensor is given by

\[
T_{\mu \nu }=\partial _{\mu }\phi \partial _{\nu }\phi -\frac{1}{2}g_{\mu
\nu }\partial ^{\alpha }\phi \partial _{\alpha }\phi -\frac{m^{2}}{2}g_{\mu
\nu }\phi ^{2}+\theta [g_{\mu \nu }\square -\nabla _{\mu }\nabla _{\nu
}+G_{\mu \nu }]\phi ^{2}\;,
\]
with $\theta =\frac{\gamma }{4}\frac{d-2}{d-1}$. The current $j^{\mu }=\sqrt{%
-g}\xi ^{\nu }T_{\;\nu }^{\mu }$ is conserved provided $\xi ^{\mu }$ is a
Killing vector. The vanishing of the energy-momentum flux density $\sqrt{-g}%
T_{0}^{i}d\Sigma _{i}$ in the asymptotic region of Eq.  (\ref{MasslessZ}) is
expressed as
\begin{equation}
\lim_{z\rightarrow 1}\frac{1}{(1-z)^{\frac{d-3}{2}}}\left( (1-4\theta
)z\partial _{z}+2\theta \frac{1}{(1-z)}\right) \phi ^{2}\rightarrow 0\;.
\label{AsymptFlux}
\end{equation}
The scalar field is given by Eq. (\ref{Scalar-Ansatz}) with $R(z)$ defined in 
Eq. (\ref{RadialHigher}). The behavior of $R(z)$ in the asymptotic region $
(z\rightarrow 1)$ is given by
\begin{equation}
R_{z\rightarrow 1}\sim (1-z)^{\beta }A[1+\mathcal{O}(1-z)]+(1-z)^{\frac{d-1}{
2}-\beta }B[1+\mathcal{O}(1-z)]\;,  \label{AsymptR}
\end{equation}
where
\[
A=\frac{\Gamma (c)\Gamma (c-a-b)}{\Gamma (c-a)\Gamma (c-b)}\;,
\]
\[
B=\frac{\Gamma (c)\Gamma (a+b-c)}{\Gamma (a)\Gamma (b)}\;,
\]
with $a$, $b$, $c$, and $\beta $ are defined in Eqs. (\ref{abcHigher}) and (
\ref{BetaHigher}), and $c-a-b=\frac{d-1}{2}-2\beta $ cannot be an integer.
Substituting the asymptotic form of $\phi $ in Eq. (\ref{AsymptFlux}), the
following condition is obtained

\begin{eqnarray}
&&AB(1-d+4\theta d)+2A^{2}(\theta -\beta +4\theta \beta )\;(1-z)^{2\beta -
\frac{d-1}{2}}+  \nonumber \\
&&B^{2}(1-d-2\theta +2\beta -8\theta \beta +4\theta d)\;(1-z)^{\frac{d-1}{2}
-2\beta }+  \label{Condition} \\
&&A^{2}C\;(1-z)^{\frac{3-d}{2}+2\beta }+B^{2}D\;(1-z)^{\frac{d+1}{2}-2\beta
}+ABE\;(1-z)  \nonumber \\
&=&0,\;  \nonumber
\end{eqnarray}
where $C$, $D$, $E$ are of the form (const $+O(1-z)$).

If the effective mass does not satisfy the bound (\ref{MT-Bound}), the
condition (\ref{Condition}) can only be satisfied if the scalar field
identically vanishes. Hence, nontrivial solutions require $\beta $ to be a
real number.

Since $\frac{d-1}{2}-2\beta $ cannot be an integer, the condition (\ref
{Condition}) can only be satisfied for $A=0$ or $B=0$. In case of $B=0$, the
condition (\ref{Condition}) is always met if
\begin{equation}
\beta >\frac{d-1}{4}\;,  \label{Cond1}
\end{equation}
which is only satisfied for the branch $\beta =\beta _{+}$. This means that
the quasinormal frequencies are found through $a|_{\beta _{+}}=-n$ or $
b|_{\beta _{+}}=-n$, where $n$ is a nonnegative integer.

If the condition

\[
\theta -\beta +4\theta \beta =0\;,
\]
holds, then Eq. (\ref{Condition}) require $\beta >\frac{d-3}{4}$, which can
also be satisfied for the branch $\beta =\beta _{-}$ in the range
\[
\frac{d-3}{4}<\beta _{-}<\frac{d-1}{4}\;,
\]
which gives rise to the bound (\ref{RangoRaroHigher}), and the quasinormal
frequencies are determined $a|_{\beta _{-}}=-n$ or $b|_{\beta _{-}}=-n$.

The case with $A=0$ is equivalent to the former, because of the 
relations
\begin{eqnarray*}
A|_{\beta _{\pm }} &=&B|_{\beta \mp }\;, \\
b|_{\beta _{\pm }} &=&(c-a)|_{\beta \mp }\;, \\
a|_{\beta _{\pm }} &=&(c-b)|_{\beta \mp }\;, \\
(\theta -\beta +4\theta \beta )|_{\beta _{\pm }} &=&(1-d-2\theta +2\beta
-8\theta \beta +4\theta d)|_{\beta \mp }\;.
\end{eqnarray*}

\end{document}